\documentclass[lettersize,journal]{IEEEtran}
\usepackage{tikz}
\usepackage{textcomp}

\newcommand\copyrighttext{%
  \footnotesize \textcopyright 2022 IEEE. Personal use of this material is permitted.
  Permission from IEEE must be obtained for all other uses, in any current or future
  media, including reprinting/republishing this material for advertising or promotional
  purposes, creating new collective works, for resale or redistribution to servers or
  lists, or reuse of any copyrighted component of this work in other works.
  DOI: 10.1109/JIOT.2023.3332507 }
\newcommand\copyrightnotice{%
\begin{tikzpicture}[remember picture,overlay]
\node[anchor=south,yshift=10pt] at (current page.south) {\fbox{\parbox{\dimexpr\textwidth-\fboxsep-\fboxrule\relax}{\copyrighttext}}};
\end{tikzpicture}%
}

\usepackage{todonotes}
\usepackage{amssymb}%
\usepackage{pifont}%
\newcommand{\cmark}{\ding{51}}%
\newcommand{\xmark}{\ding{55}}%
\newcommand{\omark}{\ding{108}}%

\usepackage{color, colortbl}
\definecolor{Gray}{gray}{0.9}

\usepackage{makecell}
\usepackage{balance}
\usepackage{listings}
\usepackage{booktabs}
\usepackage{url}
\usepackage{hyperref}
\usepackage{hyperref}
\hypersetup{colorlinks = true}
\hypersetup{linkcolor  = black}
\hypersetup{citecolor  = black}
\hypersetup{filecolor  = black}
\hypersetup{pdftitle   = IIoT Network Review}
\hypersetup{pdfauthor  = Behnke and Austad}
\hypersetup{pdfsubject = Real-Time Performance Evaluation of IIoT Communication Technologies: A review}
\hypersetup{pdfkeywords= real-time  IIoT networks TSN WLAN 5G CPS Industry4.0}
\usepackage[inline]{enumitem} %
\usepackage{tablefootnote}

\usepackage[backend=biber,
            citestyle=ieee,
            style=numeric-comp,
            sorting=none,
            ]{biblatex}
\bibliography{references.bib}

\begin{document}
\title{Real-Time Performance of Industrial IoT Communication Technologies: A Review}
\author{
\IEEEauthorblockN{Ilja Behnke\IEEEauthorrefmark{1}, Henrik Austad\IEEEauthorrefmark{2}}\\
\IEEEauthorblockA{\IEEEauthorrefmark{1}Technische Universität Berlin, Berlin. \emph{i.behnke@tu-berlin.de}}\\
\IEEEauthorblockA{\IEEEauthorrefmark{2}SINTEF Digital, Trondheim. \emph{henrik.austad@sintef.no}}
}

\maketitle
\copyrightnotice
\begin{abstract}
With the growing need for automation and the ongoing merge of OT and IT, industrial networks have to transport a high amount of heterogeneous data with mixed criticality such as control traffic, sensor data, and configuration messages.
Current advances in IT technologies furthermore enable a new set of automation scenarios under the roof of Industry 4.0 and IIoT where industrial networks now have to meet new requirements in flexibility and reliability. 
The necessary real-time guarantees will place significant demands on the networks.
In this paper, we identify IIoT use cases and infer real-time requirements along several axes before bridging the gap between real-time network technologies and the identified scenarios. We review real-time networking technologies and present peer-reviewed works from the past 5 years for industrial environments. We investigate how these can be applied to controllers, systems, and embedded devices. Finally, we discuss open challenges for real-time communication technologies to enable the identified scenarios. The review shows academic interest in the field of real-time communication technologies but also highlights a lack of a fixed set of standards important for trust in safety and reliability, especially where wireless technologies are concerned.

\end{abstract}

\begin{IEEEkeywords}
real-time, networks, IIoT, CPS, Industry 4.0
\end{IEEEkeywords}

\section{Introduction}
\label{sec:introduction}

Science and technology are bridging the mechanized industry with the digital world to automate as many processes as possible by connecting operational technologies and business networks.
Such a pervasive sensor integration combined with a high degree of network interconnectedness, both between operational technologies (OT), and between OT and "back office" systems (IT), is often described as "IT/OT convergence".
It is also known as "Industry 4.0".
Examples of envisioned Industry 4.0 scenarios are autonomous drones~\cite{lohan2018benefits}, remote control~\cite{liu2020latency}, real-time monitoring~\cite{zhan2022industrial}, and predictive maintenance~\cite{nordal2021modeling}.   

In a recent study~\cite{rwsn_industrial_AI_smart_factory_2021} regarding the usage of pervasive Industrial Internet of Things (IIoT) sensor data, a large majority of the respondents expect Industry 4.0 to provide agility to scale the production to match demands, improved flexibility to customize products, and a reduced time to market for new products.
The same participants reported that the current usage of IoT sensors in their systems was primarily used for remote monitoring of equipment, asset and material tracking, and predictive maintenance.
Hence, to fully realize the potential of Industry 4.0, pervasive sensor coverage and interconnectedness are essential.
To achieve this, great demands will be placed on the networks used in industrial systems.
What used to work for Operation Technology (OT) systems will require a much higher degree of flexibility and scalability than what has been available in the past.

Furthermore, advancements in Cyber-Physical Systems (CPSs) increase the real-time requirements for industrial networks.
CPSs combine embedded systems with cybernetic control systems to control the physical world from the networked space.
The physical processes involved often place strict real-time requirements on CPSs.
Command, control, and safety considerations of different criticality levels demand constant latency and predictability, often implemented on small and restricted embedded systems.
By attaching these embedded systems to larger networks and incorporating them into complex distributed systems, the same real-time requirements are extended to the network and its participants.
Use cases with time-critical control tasks are spread over a network, requiring predictable and reliable packet routing through it. 

However, networking concepts like the Internet Protocol (IP) and packet routing are not created to consider real-time demands. IP networks work on a best-effort basis.
Packets are indiscriminately routed over a variable number of networking devices, with very limited predictability concerning latency and arrival.
The jitter introduced by wireless communication complicates a number of envisioned scenarios including mobile devices.

To meet the real-time demands of networking introduced by envisioned IIoT scenarios, established specialized protocols and standards exist. At the same time, current works in research attend to IIoT-specific requirements such as scalability~\cite{mirani2022key}, integration of business and factory networks~\cite{hicking2021collaboration}, and mobile computing~\cite{ma2021tcda}.

In this article, we present and discuss existing standards and technologies as well as present recent works concerning real-time IIoT networking solutions. To this end, we
\begin{itemize}
    \item identify and present current IIoT scenarios attended to in research,
    \item derive their real-time networking requirements,
    \item review established real-time networking technologies,
    \item presents the most notable recent research,
    \item and discuss open challenges.
\end{itemize}

The remainder of this article is structured as follows.
Section \ref{sec:scenarios} identifies IIoT scenario classes and derives real-time networking requirements.
Section \ref{sec:background} reviews the path of networking in industrial systems and presents existing standards designed for industrial networks.
Section \ref{sec:recent} presents ongoing academic works on real-time networking.
Section \ref{sec:discussion} discusses the feasibility of existing and ongoing solutions.
Section \ref{sec:related} presents related industrial networking surveys.
Section \ref{sec:conclusion} concludes this article.

\section{Regarded IIoT Scenarios}
\label{sec:scenarios}

In this section, we perform a preliminary literature survey to identify IIoT scenarios containing real-time and network requirements. To this end, we classify the scenarios into groups as depicted in Table~\ref{tab:scenarios}.
Network and real-time requirements are derived from the scenario groups to prepare the following survey on enabling technology standards and recent works.
A total of 31 research papers with IIoT scenarios of interest have been reviewed. The papers and articles chosen were peer-reviewed and published in the past 10 years.
The extracted use cases were either the main focus of a paper or example implementations for evaluation purposes.

Using the identified scenario classes, we hope to get a better understanding of network requirements concerning predictability, latency, and bandwidth.
These requirements will then be used in the following sections to assess proposed IIoT real-time networking solutions. Furthermore, we identify relevant network components for each scenario class.
 
\begin{table*}[ht]
    \centering
    \caption{IIoT scenario classes}
    \label{tab:scenarios}
    \begin{tabular}{cl|c|c|c|c|c}
    &\thead{\textbf{Scenario Class}} & \thead{\textbf{References}} & \thead{\textbf{Predictability}\\ \textbf{Requirement}} &  \thead{\textbf{Latency} \\ \textbf{Requirement}} & \thead{\textbf{Bandwidth} \\ \textbf{Requirement}} & \thead{\textbf{Wireless} \\ \textbf{Communication}}\\ \hline
    \textbf{\textrm{I}} & \makecell[l]{Autonomous Vehicles \\ \& Mobile Robots} & \cite{lohan2018benefits, miljkovic_new_2013, singholi_review_2021, chavhan_smart_2021, lo_bello_perspective_2019,  barzegaran_fogification_2020} & high & medium & variable & essential\\ \hline
    \textbf{\textrm{II}}& Remote Control of Machines & \cite{zhohov_real-time_2018, kayan_cybersecurity_2022, kajan_control_2013, huang_sense_2019, bock2021performance} & high & high & low & not necessary\\ \hline
    \textbf{\textrm{III}}& \makecell[l]{Sensor Data Analysis\\ \& Monitoring} & \cite{kanawaday_machine_2017, blanes_80211n_2015, iqbal_cooperative_2017,  hicking2021collaboration, salhaoui2019smart} & low & low & high & not relevant\\ \hline
     \textbf{\textrm{IV}}& \makecell[l]{Worker Safety \\ \& Hazard Protection}  & \cite{kivela2018towards, zhou2017industrial, mcninch2019leveraging, reyes2013intelligent} & high & high & low & not necessary\\ \hline
     \textbf{\textrm{V}}& Task Offloading & \cite{hong_multi-hop_2019, behnke2023offloading, deng_intelligent_2021, xu_joint_2022, ma2022reliability} & medium & medium & high & not relevant\\ \hline
     \textbf{\textrm{VI}}& \makecell[l]{Industrial Wearable Systems \\ \& Augmented Reality} & \cite{kong_industrial_2019, svertoka_wearables_2021, lorenz_industrial_2018, rosales_iiot_2021, batalla_analyzing_2020, li2021helically} & low & medium & high & essential\\%
     \hline
    \end{tabular}

\end{table*}

\subsection{Scenario Classes}

\paragraph{Autonomous Vehicles \& Automation}
\label{para:sc:autonomous}
Warehouses and manufacturing sites are getting under pressure to fulfill an ever-increasing demand. To this end, conventional rigid automation is being exchanged for more flexible autonomous solutions controlled via IoT-connected sensors~\cite{chavhan_smart_2021}. Autonomously moving machines such as transport robots, forklifts, and warehouse robots pose challenges yet to be solved, especially when employed at a large scale or in close proximity to human workers. Autonomous vehicles have to navigate through dynamic environments making decisions for real-time controlled electric motors~\cite{barzegaran_fogification_2020} while being dependent on communication networks to prevent collisions and improve efficiency~\cite{singholi_review_2021, lo_bello_perspective_2019}. 

Converging these real-time and networking requirements still requires research, especially considering the necessity for wireless communication. Since moving, potentially dangerous machines are involved, predictability and latency requirements are high.
Bandwidth requirements are implementation-specific, with the main factor being whether high-volume sensor data (e.g. video, LIDAR) is transmitted for analysis. 

\paragraph{Remote Control}
Industrial Control Systems (ICS) can be changed to make use of the IIoT paradigm by physically shifting controlling entities away from the real-time device, possibly outside of the plant floor~\cite{bock2021performance}. To this end, the originally existing air gap between ICS and external networks has to be bridged~\cite{kayan_cybersecurity_2022}. Besides the resulting security implications, this also means opening up real-time control systems to conventional IP networks and protocols. Use-cases furthermore include the remote control of vehicles in inaccessible places~\cite{zhohov_real-time_2018}. 

Again, real-time requirements are to be considered as high, since we are regarding moving machines.
Some use cases furthermore need wireless communication technologies to work. Bandwidth requirements are low as only control commands need to be transmitted.

\paragraph{Data Analysis \& Monitoring} 
With the help of sensors on factory floors, a high amount of real-time data can be used to improve business automation, monitoring, and decision-making. To this end, potentially broad data streams have to be transmitted sharing a medium with real-time control traffic~\cite{blanes_80211n_2015}. Use cases include data aggregation by wireless sensor networks~\cite{iqbal_cooperative_2017} and predictive maintenance on IIoT sensor data~\cite{kanawaday_machine_2017}. IT/OT integration also plays a role in monitoring and data analysis. Real-time information from IIoT sensors can help make business decisions and enable automating certain use-cases such as active energy management~\cite{hicking2021collaboration}.

This scenario class, while somewhat dependent on real-time data streams does not contain hard real-time tasks. Hence requirements in this regard are negligible. However, due to the large amount of collective data produced by the high number of sensors, bandwidth requirements are usually high. With data-intensive applications benefitting from offloading, bandwidth requirements can be considered high for many use cases.

\paragraph{Worker Safety \& Hazard Protection}
Most considered works regard the introduction of IIoT to safety critical systems as a relevant risk factor. Safety in this case means the physical protection of workers and equipment.
Due to the attack vector granted by the new connectivity, safety is more conjoined with security than ever~\cite{kivela2018towards}. Yet, some works promote the use of interconnected devices for better safety in high-risk environments such as in the mining industry. IIoT technologies can be used to monitor access to hazardous areas around operating machinery, improve documentation/monitoring of maintenance that requires shutdown of the machinery, and prevent unexpected startup or movement during machine maintenance activities~\cite{mcninch2019leveraging, zhou2017industrial, reyes2013intelligent}. 

Safety solutions distinguish between informational and automatizing setups. Purely informational scenarios react to real-time sensor data by issuing alarms or notifying workers and human operators. Smart monitoring and data analysis in the cloud make these scenarios and their real-time requirements similar to the \textit{Data Analysis} class. Systems that automatically react to hazards such as fire, toxic gases, or dangerous proximity require the highest predictability and latency. However, reactive systems should work independently, making distributed solutions unsuitable.

\paragraph{Task Offloading} 
IIoT devices, like traditional embedded systems, only have limited computing capabilities due to power and space constraints.
To this end, current research, especially in the field of mobile edge computing, considers the offloading of delay-sensitive tasks to local servers.
The main challenges are the guarantee of maximum delays over shared networking channels and compute resources~\cite{deng_intelligent_2021}.
Subject to offloading is mostly soft real-time tasks for better Quality of Service (QoS) where a balance between latency and reliability has to be found~\cite{ma2022reliability}.

Due to the nature of offloading, hard real-time tasks with high requirements in predictability are not considered.
At the same time, while offloading to more powerful machines can reduce computation time, latency requirements that are too tight to accommodate network delays cannot be accommodated.

\paragraph{Wearable Systems \& Augmented Reality}
Industrial wearable systems are the means of human-machine interaction (HMI) for most of the other listed scenario classes. With manufacturing, maintenance, and control still requiring human interaction, wearables present the necessary interface to embedding human workers in smart factories. Wearable systems enable remote control~\cite{li2021helically}, real-time monitoring via augmented reality~\cite{batalla_analyzing_2020}, sensor data aggregation~\cite{rosales_iiot_2021}, and safety systems~\cite{svertoka_wearables_2021} while adding the requirements of energy efficiency, size (weight), and wireless networking. Examples roughly follow two streams: human interaction devices and data collection devices~\cite{kong_industrial_2019}. For this scenario class, only the network requirements specific to wearables and augmented reality are considered. If the devices are used for i.e. remote control, Scenario Class II covers the requirements.

\subsection{Deriving Network Requirements}
To derive network requirements for the scenario classes, we need to somewhat generalize the regarded use cases. In the end, it comes down to the specific requirements of the underlying real-time tasks and how messages (i.e. network packets) are integrated into the control flow.
The requirements broadly quantified in Table~\ref{tab:scenarios} therefore only give a rough overview of the focal points and limiting factors related to the classes.
The spectrum of feasible IIoT scenarios increases with the strictness of real-time guarantees networks can adhere to while not limiting their flexibility. 
Furthermore, wireless communication technologies that can fulfill these guarantees are the necessary enablers for many of the identified scenarios.
Interfacing general IP networking into real-time systems is necessary for use cases requiring external access or cloud computing.
However, for most real-time relevant scenarios IT/OT integration is not the enabler. Real-time scenarios become more complex the further away participants are, limiting real-time guarantees and criticality levels. 

\section{Industrial Networking}
\label{sec:background}

The introduction of the first Programmable Logic Controller (PLC), Modicon model 084 in 1969, changed how production systems were designed, implemented, operated, and maintained. 
Improving mechanical systems and machinery with relay-based control loops was but a small step compared to what PLCs brought to the industry.
Whole cabinets of relays could be replaced with a relatively small number of compact devices saving both space and power.
Where the first PLC only had 16 inputs, 16 outputs, and a meager 1 kB of memory, contemporary PLCs have I/O in the order of hundreds, abundant memory, and much higher speeds.

The next sections will cover different industrial networks and common requirements.
Table \ref{tbl:network_tech} presents the connection between networks and use cases.

\subsection{Fieldbuses}
With improved industrial automation, communication became an obvious hurdle to tackle.
In 1979, Modicon developed and published the serial communication protocol Modbus, an application-layer protocol.
Shortly thereafter the work to standardize Profibus started in Germany and was finally published in 1989.
These first communication protocols would be the inaugural Operational Technology (OT) networks.

As the first fieldbus protocols grew in popularity, so did issues concerning interoperability.
The different protocols were seldom compatible, and few vendors could support multiple protocols.
After a somewhat turbulent era often referred to as "the Fieldbus War"~\cite{felser_fieldbus_2002}, an agreement was made in the late 1990s to create a common standard.
When the first fieldbus standard (IEC 61158) was published, the final product was essentially a collection of \textit{all} the competing standards at the time.
The latest revision of IEC61158 \cite{iec61158} lists 26 different fieldbus protocols grouped into services and protocols.
Since then, it has become apparent that the industry is increasingly looking towards open standards \cite{kayan_cybersecurity_2022} to avoid vendor lock-in and a more transparent way to handle security vulnerabilities to name a few.

In this same decade, the Controller Area Network (CAN) was also standardized by the automotive industry.
CAN is a serial communication protocol where the message identifier also serves as the message priority.
The protocol is optimized for short messages and has the priority encoded in the message identifier.
This provides a non-destructive address arbitration which enables messages to pass without any delay induced by lower priority messages.

\subsection{Packet Switched Networks}
At the same time, the IT sector underwent a networking revolution of its own where Ethernet would ultimately end up as the de-facto standard.
After its initial version was published in 1985, it has since moved the throttle from 10 Mbps to 400 Gbps~\cite{ieee:802.3-2022} with its sight firmly set on 800 Gbps.
Data is split into discrete packets, each identified using a 48-bit Media Access Control (MAC) address giving ample room for growth.
To connect hosts, systems can be chained together to form a ring topology or by using ethernet bridges (where store-and-forward switches are the most common).
Each frame is individually routed, and the network can quickly adapt to changes in topology and assign new routes to frames while in transit.
 
In 1974, Cerf et al. published TCP/IP \cite{rfc675} which was to become the backbone of the "network of networks" lovingly called "the Internet".
TCP is a connection-oriented protocol that when traffic moves in one direction, control traffic will move both ways allowing TCP to resend lost data and adjust the rate to the slowest link along the path.
UDP is the connectionless sibling of TCP without any control traffic and thus offers no indication that traffic arrives at the destination.

\subsubsection{Industrial Ethernet}
Industrial Ethernet encompasses the usage of Ethernet in industrial settings.
The target applications typically have both latency and reliability requirements, which drives the design of protocols away from traditional Ethernet approaches for collision detection and avoidance.
The most common real-time Ethernet protocols are EtherNet/IP\footnote{IP is for
  \textbf{Industrial} Protocol.}, Profinet, and EtherCAT~\cite{watteyne_industrial_2016}.
PROFInet is a translation of PROFIbus to run over an Ethernet network.
EtherCAT is often used in industrial control and automation due to its speed and determinism.
TTEthernet is a congestion-free network based on Ethernet that provides Time Triggered service for critical traffic, Rate Constrained for event-triggered traffic as well as a Best-Effort (BE) service.
EtherNet/IP is an adaption of DeviceNet to Ethernet, it uses the Common Industrial Protocol (CIP) over TCP and UDP.

\subsubsection{Time Sensitive Networking}(TSN) is a series of IEEE Standards for switched Ethernet~\cite{ieee:802.1ba_2021, ieee:802.1Q-2018} where Commercial Off-the-Shelf (COTS) networks can be configured to give bounded latency and extremely low packet loss for critical traffic on the data link layer.
TSN removes some of the initial robustness of Ethernet by requiring static routes but gains lower jitter and less out-of-order delivery.
In addition, TSN defines strict and fine-grained Quality of Service (QoS) mechanisms which are covered in more detail in Section \ref{sec:psn_qos}.
\subsection{Wireless networks}
Wireless sensors are cheaper, easier to install and maintain than their wired counterparts and bear the promise of an infrastructure that is vastly more scalable.
In some scenarios, the cost-savings can be as high as 60-90\%~\cite{power_reduce_2009,8558500} compared to wired solutions.
It can even be the only viable option (e.g., the Tire Pressure Monitoring System (TPMS), tool deflection measurements~\cite{ostling_cutting_2018}).

The downside of wireless networks is the shared medium; the integrity of the network is fully dependent upon cooperative participants that only transmit during allocated transmit slots.
Frequency bands needed for wireless protocols are a tightly managed resource, and only a few bands are available for free use, most notably in the 2.4GHz and 5 GHz range.
Where a wired network can be fairly resilient to signal interference, have high and predictable bandwidth, and require an adversary to be close to the wired infrastructure to eavesdrop, wireless systems have no such luxury.

Over the years, many wireless protocols have been defined with different characteristics such as high bandwidth, low latency, long-range, a high number of addressable hosts, and robust resistance to interference.
Oftentimes, these attributes will be at odds; i.e. with longer ranges come lower bandwidth and higher latency, and high bandwidth can make the traffic more susceptible to interference due to the denser encoding.

Both WirelessHART and ISA 100.11a are common industrial protocols that are based on IEEE 802.15.4 and operate in the unlicensed 2.4GHz frequency band.
ZigBee, while also being based on 802.15.4, can be found in some industrial settings but is primarily intended for home automation and low-criticality systems.
Bluetooth is mentioned due to its pervasiveness in personal handheld devices.
With its low power, it is only capable of transmitting data over a few meters and is primarily intended for short-range personal area networks.
The upside is rather low power consumption ample access to complete modules, and comparably high available bandwidth.

In recent years, Wireless LAN (WLAN) has become a common technology in most households, but the technology has not been reliable or deterministic enough for industrial settings.
With the recent WLAN6(E), this has improved markedly, and with the availability of Multiple-Input/Multiple-Output (MIMO), access is more reliable and less prone to interference from other senders.
The latest version, WLAN6E has also markedly improved the clock accuracy for protocols such as PTP, which makes sensor aggregation more accurate.
In addition to the rather crowded 2.4GHz band, WLAN can also operate in the 5GHz and 6GHz range allowing for higher bandwidth at the cost of higher signal attenuation from obstacles along the path.

The fifth generation of mobile telephony (5G) from the 3GPP aims to cover all areas of wireless communications from cellular networks, IoT devices, and industrial networks.
5G is designed to operate in licensed bands where a site license specifying both assigned frequency range and spatial location is required.
Interference from other networks should therefore be at a minimum.
Three broad use cases have been defined for 5G: cellular data (\emph{enhanced Mobile Broadband, eMMB)}, IoT (\emph{massive Machine Type Communications, mMTC)} and low-latency, critical traffic (\emph{Ultra-Reliable, Low Latency, URLLC)}. 
It is important to note that these use cases are not introduced in any single release, but support for each is added as increments with different capabilities in each release.
With the first 5G release from 3GPP, Release 15, support for New Radio, rudimentary slicing, and the IoT profile from 4G were among the many parts included.
The next release saw the introduction of improved slicing ("VLAN for wireless networks") and redundant transmission for high-reliability communications.
Finally, with the latest approved release (Rel17, March 2022), improvements in backhaul networks, RAN slicing, public safety, and non-public networks (NPN) have been ratified.
NPN is often called ``private 5G`` and makes it possible to run a 5G network as a standalone network as a local service rather than a public mobile operator.
Running a 5G NPN is highly relevant in industrial automation as slicing, URLLC and access to licensed radio bands can greatly aid in security, reliability, and latency.
Whereas eMMB is the use case most commonly supported by public providers, URLLC will be most relevant over NPN.
Although the specification seems impressive, not much hard data is currently available to evaluate the performance of URLLC.

\begin{table*}[h]
\centering
\caption{Technologies along Security, Determinism, and Bandwidth. Legend: \cmark: well suited, \xmark: not well suited}
\label{tbl:network_tech}
\begin{tabular}{r | l l r r | c c c c c c}
 & & & & & \multicolumn{6}{c}{\textbf{\textit{Relevant Scenarios}}}\\
    & \textbf{Security}
    & \textbf{Determinism}
    & \textbf{Bandwidth} &
    & \textbf{\textrm{I}} & \textbf{\textrm{II}} & \textbf{\textrm{III}} & \textbf{\textrm{IV}} & \textbf{\textrm{V}} & \textbf{\textrm{VI}}\\
\hline
\textit{Fieldbuses} &&&&&\\
CAN         & E & 10 ms &  1 Mbps & \cite{godoyDesignCANbasedDistributed2010, rubiobenitoPerformanceEvaluationFour1999} & %
\xmark & \cmark & \xmark &\cmark & \xmark & \xmark \\
CAN-FD      & E & 100 $\mu$s &  5 Mbps & \cite{austermannConceptsBitrateEnhancement2020} & %
\xmark & \cmark & \xmark & \cmark & \xmark & \xmark\\
Modbus      & E & 10 ms & 115 kbps & \cite{rubiobenitoPerformanceEvaluationFour1999} & %
\xmark & \cmark & \xmark & \cmark & \xmark & \xmark\\
PROFIBUS    & E & 10 ms & 12 Mbps & \cite{junAnalysisPROFIBUSDPNetwork2005, rubiobenitoPerformanceEvaluationFour1999} & %
\xmark & \cmark & (\cmark) & \cmark & (\cmark) & \xmark\\
\midrule
\textit{Wired network}&&&&&\\
Modbus TCP  & E & 45$\mu$s (ideal conditions) & 100 Mbps & \cite{modbus_smartgrid} & %
\xmark & (\cmark) & \cmark & \cmark & \cmark & \xmark\\
PROFINET    & E     & 10-100 ms &  1 Gbps & \cite{profinet_chans} & %
\xmark & (\cmark) & \cmark & (\cmark) & \cmark & \xmark\\
EtherNet/IP & E   & 1-2 ms &  1 Gbps & \cite{alessandriaPERFORMANCEANALYSISETHERNET2007} & %
\xmark & \cmark & \cmark & \cmark & \cmark & \xmark %
\\
TSN CBS     & E,A   & 2 / 50 ms & 40 Gbps & \cite{zhao2017timing} & %
\xmark & (\cmark) & \cmark & \cmark & \cmark & \xmark\\
EtherCAT    & E   & 34 $\mu$s & 10 Gbps & \cite{nguyenEtherCATNetworkLatency2016} & %
\xmark & \cmark & \xmark & \cmark & (\cmark) & \xmark
\\
TTEthernet  & E,A    &  low, $\mu$s (offline schedule) & 1 Gbps & \cite{zhao2017timing} & %
\xmark & \cmark& \cmark & \cmark & (\cmark) & \xmark\\
TSN ST      & E,A   & 100 $\mu$s & 40 Gbps &\cite{zhao2017timing} & %
\xmark & \cmark & (\cmark) & \cmark & (\cmark) & \xmark \\
\midrule
\textit{Wireless network}&&&&&\\
WirelessHART & I,C,A & 10 ms  & 250 kbps & \cite{godoyEvaluatingSerialZigBee2012} &%
\cmark & (\cmark) & \xmark & (\cmark) & \cmark & \xmark\\
ISA 100.11a  & I,C,A & 10 ms  & 250 kbps & \cite{godoyEvaluatingSerialZigBee2012} &%
\cmark & (\cmark) & \xmark & (\cmark) & \cmark & \xmark\\
Bluetooth 5.0& I,C,A & 10-100 ms & 48 Mbps & \cite{rondonEvaluatingBluetoothLow2017} & %
\xmark & \xmark & \cmark & \xmark & (\xmark) & (\cmark)\\
WLAN 6       & I,C,A & 1 ms (ideal conditions) & 9.6 Gbps & \cite{bankovEnablingRealtimeApplications2019, oughton_revisiting_2021} &%
(\cmark) & \xmark & \cmark & \xmark & (\cmark) & \cmark\\
5G eMMB     & I,C,A & 10-100 ms  & 10 Gbps & \cite{oughton_revisiting_2021} & %
\xmark & \xmark & \cmark & \xmark & (\cmark) & (\cmark)\\
\hline
\end{tabular}
\end{table*}

\subsection{Quality of Service}
\label{sec:psn_qos}
The key strength of Ethernet is its ability to seamlessly accept new systems, adapt to topological changes, and handle variable amounts of traffic.
For industrial networks where rigid deadlines are the norm rather than the exception, this dynamism can jeopardize transmission delays and be a major problem.
First, packets can take one of multiple \emph{routes} in such a network leading to different arrival orders, variance in delay, or being completely lost.
Secondly, once a packet has started transmitting, no other packet, regardless of importance can be transmitted.
Since most network bridges are \textit{store and forward}, the entire packet must be fully received before it can be forwarded to the next stop.
This requires buffer capacity before newly arrived packets can be forwarded to the next stop.
During times of high traffic, these buffers can be exhausted and the bridge will have no other choice than to drop packets.

\paragraph{Standard PSN QoS measures}
QoS describes the treatment of frames belonging to a particular class (or stream) compared to other frames.
In this context, a stream, or a flow, is a set of packets that logically belong together and should be treated similarly by the network.
Several measures are available, yet whereas these are highly relevant to IT networks and the Internet, fewer apply to the needs of industrial networks.

Differential Services (\emph{DiffServ}) is a highly scalable class-based service where each stream is assigned to a particular service level and it is up to the network administrator to assign resources to each level.
DiffServ is intended to be used across large networks and networks of networks where there can be more than one operator.
However, the actual service provided to a class may differ between operators.
What is more, there is no way to differentiate between streams within a class, meaning as the network grows, the interference from other streams increases.

Integrated Services (\emph{IntServ}) maintains a 1:1 mapping between streams and granted service levels.
This ensures consistent QoS throughout the network.
Since IntServ is stream-based, its scalability is tied directly to the number of streams.
For large networks or networks of networks, the number of active streams will quickly outgrow the available resources in each bridge.

For industrial control applications, the uncertainties of DiffServ may not suffice, and for large-scale sensor networks as foreseen in Industry 4.0, IntServ will have scalability problems.
An effective solution demonstrated by Harju and Kivimaki~\cite{harju_co_operation_2000} is to use IntServ on the edges to shape and limit all incoming streams, and DiffServ in the core network to handle the bulk of traffic.
This hybrid approach can deliver adequate services, especially for large networks, which makes this one of the possible approaches presented by the Deterministic Networking (DetNet) group of IETF.

\paragraph{Time Sensitive Networking}
TSN works by reserving buffer capacity for incoming streams and using different traffic queues on the egress port.
An outgoing queue can have a shaper attached to it, which changes the traffic pattern to conform to the desired behavior.
The Credit Based Shaper (CBS, ~\cite{ieee:802.1Qat}) is a \emph{class based} shaper that will group streams for a given traffic class into a single queue.
Traffic belonging to this queue is then shaped so that it does not exceed the reserved bandwidth over a short service interval.
Since \emph{every} bridge in the domain is required to support these features, the edges will effectively limit the inflow, and the core bridges will work to reduce bursts.
CBS is specifically designed to eliminate transient network overloads from bursty traffic and this has been proven to work using Network Calculus~\cite{azua_avb_modelling_2014, nc_intro_tsn_maile_2020}.

With the Time Aware Shaper (TAS~\cite{ieee:802.1Qbv-2015}), TSN provides Time Division Multiplexing (TDM) by aligning transmission windows along the path of the traffic.
For scheduled traffic (ST) to work, all bridges must be part of a tightly synchronized time domain.
When configured properly, TAS can yield extremely low delay variations and give an upper bound of 100 $\mu$s transmission delay.

However, even with minor time errors, carefully adjusted transmit windows can become desynchronized and effectively delay scheduled traffic for a complete TDM cycle.
As the network complexity grows, this problem only worsens.
This makes TAS a complex and difficult scheduler to operate and thus suitable for only the most critical streams.
It also requires the network to be centrally managed.
To allow more sporadic, yet critical traffic and reduce the operational complexity at the same time, a new urgency-based scheduler has been adopted by the working group.
The Asynchronous Traffic Shaper~\cite{specht2016urgency} has per-class queues and per-stream shaping and uses its internal clock and allows for mixing traffic types such that it can handle sporadic, critical events whilst enforcing rate limits to reduce the impact of bursts.
It has been shown that ATS will not increase the worst-case delay~\cite{nasrallah_ultra-low_2019} of traffic through the network.

CBS is appropriate for traffic that should be regular but occasionally bursty, whereas TAS is best suited for critical traffic that is regular and must be expedited through the network.
For important, yet sporadic or aperiodic traffic (e.g. monitor alarms, events), ATS is appropriate.

It is worth noting that the reservation of streams may fail if one or more bridges are unable to accommodate the requirements and that can only reliably be detected at run-time.
It is possible to discover this \emph{a priori} to a certain extent (Maile et al.~\cite{nc_intro_tsn_maile_2020}), but as small changes to the network can result in new routes for traffic, previous scenarios may no longer be possible.
This makes it difficult to realize the full potential of a network using a fully distributed model where each node announces or reserves resources for a stream individually.
Instead. a centralized controller can handle end-station interaction and bridge configurations.
The \emph{Centralized User Configuration} (CUC) and \emph{Centralized Network Configuration} (CNC) provide single points of contact for end-stations.
New streams are evaluated and by using the total overview, an optimal path can be created and forwarded to the CNC, which in turn configures all the required bridges before the end stations are notified and traffic can begin.
For TAS, such a centralized controller is a requirement.

\subsection{Common Network Requirements for Industrial Automation}
\label{sec:rt_net_req}
Devan et. al \cite{devan_survey_2021} list 4 criteria as fundamental for a wireless industrial network: 1) Secure against malicious intruders and misconfigured devices, 2) easy and dependable access to sensor data, 3) Interoperability between vendors and protocols, and 4) active research community to adapt and grow to future needs.
Missing in this list is the need for deterministic transport.
From the discussion in Section \ref{sec:psn_qos}, we can extend this list to also include wired industrial networks and include bounded latency and time synchronization as relevant metrics.

An updated requirement is thus:
\begin{itemize}
\item \textbf{Open Standards} Available standards and royalty-free technology should be the preferred approach as this will allow for better interoperability between vendors and a more open community that encourages academic research.
\item \textbf{Security} Not only should the content be shielded from external eyes, but the network should be able to detect if traffic has been altered, delayed, or injected into the network by a third party.
\item \textbf{Reliability and Availability} Network delivery must be reliable in that once sent, the sender should be confident that it will arrive at the destination within a known time frame. At the same time, a sender should expect the network to be available to accept new data within the pre-defined constraints (not overstepping BW bounds, etc.). The network should also be robust against jamming and other malicious attempts to disrupt the service~\cite{pirayesh_jamming_2022}
\item \textbf{Latency} Especially closed control loops are sensitive to delay variations, but for any streams, expecting delivery within a certain time frame is a requirement.
\item \textbf{Time synchronization} As networks are used to build distributed systems, a shared understanding of time (what is often called a shared time domain) is needed. The accuracy of the domain is dependent upon how well the network can forward time synchronization messages.
\end{itemize}

In Table \ref{tbl:network_tech}, the most common solutions found in industrial networks are listed.
Each is briefly evaluated along 3 common axes:
\begin{enumerate*}
\item Security,
\item Determinism (i.e. jitter), and
\item Bandwidth (upper limit)
\end{enumerate*}.
For brevity, Reliability, Availability, and Latency have been combined into ``Determinism`` as all the aforementioned requirements will directly affect the determinism of the traffic.
Security can be further divided into
\begin{enumerate*}[label=\emph{\roman*)}]
    \item \textbf{E}xternal: the protocol relies on external security measures,
    \item \textbf{I}ntegrity: messages are accompanied by robust checksums to detect modification,
    \item \textbf{C}onfidentiality: each message is encrypted using a secret known only to authorized parties,
    \item Secure \textbf{A}uthentication and Authorization: each node on the system must be securely identified before being granted access to the network
\end{enumerate*}
\footnote{Both wired and wireless Ethernet can achieve better security by including standards such as Port-based authentication (802.1X) and Secure Device Identity (802.1AR).}
In addition, each row is weighted against the list of use cases in Table \ref{tab:scenarios} with \cmark indicating a good fit, \xmark ~a poor.
A (\cmark) indicates that the protocol can support it under ideal conditions, but it is not well suited.
Likewise, (\xmark) indicates that the protocol \emph{can} be made to work, but it is probably not a great idea.

\section{Ongoing Efforts towards Real-Time Networks}
\label{sec:recent}
When it comes to the inherent predictability requirements of real-time applications, packet-switched networking alone is not adequate.
This is because the load on individual network hardware in larger networks is unpredictable, and packet transmission times (as well as routes) are variable.
In addition, the packet loss rates of wireless transmissions are always difficult to account for.
As described in the previous section, there are industry standards that address these issues, either by sticking to packet-less bus communication or by adding specific real-time capabilities to Ethernet.
However, with the convergence of IT and OT networks, as well as the highly mobile nature of the IIoT scenarios mentioned above, further real-time networking solutions need to be found.

In this section, we present recent work on real-time IIoT networking. We consider peer-reviewed articles and papers from the last 5 years. The works can be either novel techniques, adaptations, or evaluations of existing standards optimized for real-time IIoT scenarios. The reviewed works are divided into subsections of equal levels of abstraction and underlying technologies.
Table~\ref{tab:comparison_recent} compares the works qualitatively concerning their scope and applicability to our scenario classes. Entries marked with \cmark or \xmark ~determine whether the work is feasible for the scenario class.
Entries marked with \omark ~are works that do not directly facilitate a scenario class but also do not prohibit it.
It also indicates the type of source institution (academic or industry) and whether the work supports wireless communication.

\begin{table*}
    \centering
    \caption{Recent works}
    \label{tab:comparison_recent}
    \begin{tabular}{c|c|c|c|c|c|c|c|c|c|c} 
       \textbf{Work} & \textbf{Scope} & \textbf{Issue(s) addressed} & \textbf{Source} &\textbf{\textrm{I}}&\textbf{\textrm{II}}&\textbf{\textrm{III}}&\textbf{\textrm{IV}}&\textbf{\textrm{V}} & \textbf{\textrm{VI}} & \textbf{Wireless} \\ \hline
         \cite{ishtaique_ul_huque_system_2019} & SDN architecture & IT/OT; flexibility & academia &\omark&\xmark&\cmark&\xmark&\cmark&\omark & \cmark\\ %
         \cite{kiangala_effective_2021}  & IP architecture & reliability \& determinism & academia & \xmark&\cmark&\omark&\cmark & \cmark & \xmark & \xmark \\ %
        \cite{foschini_sdn-enabled_2021} & SDN architecture & IT/OT; security & academia &\xmark&\xmark&\cmark&\xmark&\cmark&\xmark & \xmark\\ %
        \cite{li_practical_2020} & OPC UA & TSN integration & academia & \xmark & \cmark & \cmark & \omark & \omark & \xmark  & \xmark\\ %
        \cite{morato_assessment_2021} & OPC UA & TSN assessment & academia & \xmark & \cmark & \cmark & \omark & \omark & \xmark & \cmark \\ %
        
        \rowcolor{Gray}
        \cite{wang_adaptive_2018} & SDN for IIoT & offloading & academia &\xmark&\omark&\cmark&\xmark&\cmark&\xmark & \cmark \\
        \rowcolor{Gray}
        \cite{lin_dte-sdn_2018} & SDN for traffic engineering & delay sensitivity & academia &\xmark&\cmark&\xmark&\cmark&\omark&\xmark & \xmark\\ %
        \rowcolor{Gray}
        \cite{zeng2019time} & SDN industrial Ethernet & flexibility; determinism; scale & academia &\xmark&\cmark&\cmark&\omark&\cmark&\xmark  & \xmark\\ %
        
        \cite{schriegel_migration_2021} & Time-aware forwarding & Profinet/TSN migration & academia &\xmark&\cmark&\cmark&\cmark&\omark&\xmark  & \xmark\\ %
        \cite{bujosa2022hermes} & TSN scheduler & scalability & academia &\xmark&\omark&\cmark&\omark&\cmark&\xmark  & \xmark\\  %
        \cite{chaine2022egress} & TSN traffic shaping & scalability; complexity & aviation &\xmark&\omark&\cmark&\xmark&\cmark&\xmark & \xmark\\ %
        \cite{bruckner2019introduction} & OPC UA with TSN & network infrastructure&automation&\xmark&\cmark&\cmark&\cmark&\cmark&\xmark  & \xmark\\ %
        \cite{8823854}& TSN using DDS & heterogeneous TSN; flexibility & academia & \xmark&\omark&\cmark&\cmark&\xmark&\xmark  & \cmark\\ %

        \rowcolor{Gray}
        \cite{cavalcanti2019extending} & Wireless TSN & determinism; reliability & chip manufacturer &\cmark&\cmark&\cmark&\omark&\xmark&\cmark  & \cmark\\ %
        \rowcolor{Gray}
        \cite{sudhakaran_enabling_2021} & Link layer/application mapping& wireless TSN applicability & chip manufacturer &\cmark&\cmark&\cmark&\omark&\xmark&\cmark  & \cmark\\ %
        \rowcolor{Gray}
        \cite{yun_rt-wifi_2022} & RT-WiFi over SDR & determinism; flexibility & academia&\cmark&\cmark&\cmark&\omark&\xmark&\cmark  & \cmark\\ %
        
        \cite{walia_5g_2019} & 5G slicing management & determinism; isolation & telecom &\omark&\omark&\omark&\cmark&\cmark&\cmark  & \cmark\\ %
        \cite{aijaz_private_2020} & 5G private networks & assessment & telecom &\omark&\omark&\omark&\omark&\omark&\omark  & \cmark\\ %
        \cite{wen_private_2022} & 5G private networks & assessment & telecom/car &\omark&\omark&\omark&\omark&\omark&\omark  & \cmark\\ %
        
        \rowcolor{Gray}
        \cite{behnke2023towards} & RT-aware packet reception & determinism & academia &\xmark&\cmark&\cmark&\omark&\omark&\xmark  & \cmark\\ %
        \rowcolor{Gray}
        \cite{behnke_priority-aware_2022} & RT-aware NIC & determinism &  academia&\xmark&\cmark&\cmark&\omark&\omark&\xmark  & \cmark\\ %
        \rowcolor{Gray}
        \cite{johansson2022priority} & RT-aware MAC filtering & determinism & academia &\xmark&\cmark&\cmark&\omark&\omark&\xmark  & \cmark\\ %
        \rowcolor{Gray}
        \cite{austad_ftc}&Timed C extension & TSN-aware applications & academia & \xmark & \cmark &\cmark&\cmark&\omark&\xmark  & \xmark\\ %
        \hline
    \end{tabular}

\end{table*}

\subsection{Network Architectures}
The field of research with the broadest scope regards network architectures for real-time IIoT environments.
Next to architectures realized by SDN, this segment also includes research on IT/OT integration and OPC UA.

In~\cite{ishtaique_ul_huque_system_2019} Ishtaique ul Huque et al. present a system architecture for time-sensitive heterogeneous wireless distributed software-defined networks. They derive this system architecture from enabling state-of-the-art technologies and their requirements.
Besides the manageability of parameters in heterogeneous networks, the authors hope to facilitate TSN integration with legacy networks. 

Kiangala and Wang combine zero-loss redundancy protocols, TSN, and edge computing concepts to realize an intra-domain network architecture attending to the reliability and predictability needs of time-critical IIoT applications~\cite{kiangala_effective_2021}.
Their solution is entirely based on IP networking.
Instead of manageability via SDN they focus on technological implementations for high reliability and determinism. 

Network architectures have been a focus in recent works primarily due to the IT/OT convergence requirements of Industry 4.0 and IIoT applications. Foschini et al. present an SDN-enabled architecture for this and analyze its behavior during DDoS attacks~\cite{foschini_sdn-enabled_2021}.
With a multi-layered network architecture, they aim to provide layer-specific security and real-time properties while making data from IIoT devices usable inside the business network of an industrial plant with autonomous vehicles.
Within their DDoS attack, they showed that mere changes to the network architecture cannot secure the highly critical lower levels of the network (i.e. physical machines).

The Open Platform Communication Unified Architecture (OPC UA) was introduced in 2008 as an interoperability standard for reliable information exchange and has gained momentum with discussions on IIoT networks. In recent years Li et al. presented an OPC UA architecture with TSN capabilities for the IIoT~\cite{li_practical_2020} and Morato et al. published a general assessment of OPC UA implementations under current Industry 4.0 requirements~\cite{morato_assessment_2021}. Both works showed that OPC UA can be used to integrate real-time communication requirements into Ethernet-based networks in the manufacturing industry.

\subsection{SDN}
When industrial networks become more complex and contain mixed computing systems and real-time requirements Software-Defined Networking can help with managing the resulting heterogeneity.
Wang and Li present an SDN-based solution for task offloading in IIoT environments \cite{wang_adaptive_2018}. A computing mode selector is implemented in the SDN controller and offloaded tasks are given priority based on real-time parameters. The network transmits offloaded tasks to fog computing resources in order of this priority. 

Another solution that addresses mixed timing criticality in large networks is proposed by Lin et al. \cite{lin_dte-sdn_2018}. SDN is used to implement a traffic engineering approach to schedule the transfer of delay-sensitive traffic. This works using real-time analysis of the network and monitoring the QoS metrics of participating links. Based on these metrics a dynamic scheduler with multi-path routing capabilities manages the flows.

Zeng et al. address flaws of Industrial Ethernet, namely poor scalability, insufficient self-configuration capabilities, and increased costs due to the use of proprietary hardware and present a time-slotted software defined Industrial Ethernet \cite{zeng2019time}. The approach contains a time synchronization mechanism based on PTP and a system architecture for time slot-based industrial switches. Using SDN, the implementation becomes scalable, and reconfigurable, and is not dependent on frequent infrastructure changes.

\subsection{TSN}
In recent years, researchers have addressed various shortcomings of TSN by extending, improving, and integrating it with other systems and frameworks.
Schriegel and Jasperneite worked on a bridging mode called Time-Aware Forwarding (TAF) \cite{schriegel_migration_2021} to increase the flexibility of TSN and thereby accelerate the migration from Profinet to TSN in industrial communication.
With the approach, already existing networking hardware can be made compatible with TSN.  

Tackling the scheduling of time-triggered traffic in TSN, Bujosa et al. present Hermes, a heuristic multi-queue scheduler \cite{bujosa2022hermes}.
With this improvement to the synthesis algorithm of gate control lists in TSN networks, they increase network scalability by reducing its scheduling complexity. 

To control network jitter, incoming packets are buffered along the routing path in a TSN network. Chaine et al. propose a solution \emph{Egress TT} that performs this buffering only at the final network node \cite{chaine2022egress} and optimizes it for network scalability.
The solution allows to use of non-TSN networking on the routing path and reduces computational complexity while somewhat increasing latencies. 

In~\cite{bruckner2019introduction}, Bruckner et al. showed how the ubiquitous OPC-UA protocol benefits from using TSN.
Similarly, in~\cite{8823854} Agarwal et al. demonstrated improved reliability and reduced delay for Data Distribution Services (DDS) when TSN was used to protect and expedite the traffic through the network.

\subsection{Real-Time Aware Wireless Technologies}
The need for reliable and deterministic wireless communication can be derived from most scenarios presented in Section \ref{sec:scenarios}. The following works from the past years attend to real-time wireless communication. 

A seemingly natural step is to transfer TSN methods to work with wireless technologies.
The concomitant challenges and research objectives have been thoroughly investigated by Cavalcanti et al.~\cite{cavalcanti2019extending}.
Next to a comprehensive overview of state-of-the-art wireless technologies and IEEE 802.1 TSN standards, they discuss new approaches on top of next-generation wireless technologies (e.g. WiFi 6, WigGig, and 5G) to overcome the radio-inherent challenges concerning packet loss and jitter. 

Sudhakran et al. build on top of this work and present a methodology to map application layer timing requirements in a collaborative robot application to the link layer, based on wireless TSN over WiFi~\cite{sudhakaran_enabling_2021}.
Rather than going further into the TSN standards, they address how to classify traffic, extract time-critical flow parameters, and define an efficient schedule for the end-to-end QoS approach.

Yun et al. recognize the older real-time aware WiFi (RT-WiFi~\cite{wei2013rt}) technology which is implemented on common off-the-shelf hardware and hence hard to update and maintain.
In their work, they propose a software-defined radio approach of RT-WiFi implemented on an FPGA called SRT-WiFi ~\cite{yun_rt-wifi_2022}. They provide a fully open system architecture and perform extensive evaluations on a multi-cluster SRT-WiFi testbed.

\subsection{Fifth Generation Mobile Networks} 
The fifth generation of mobile networking (5G) promises several improvements and enabling technologies for Industry 4.0 and smart factory applications. Among them are network slicing, private management, and ultra-low latencies.

Network slicing allows for logically separated virtual networks over the same physical 5G network. Both, the integration of different business layers as well as connection-specific latency control could be realized with this.
Walia et al. studied the usage of network slicing in a smart factory and developed a management model for 5G slicing in the domain~\cite{walia_5g_2019}. 
As the effective usage of 5G for Industry 4.0 environments requires self-managed or "private" 5G networks, some recent works have focused on their implementation and feasibility.
Aijaz et al. present an overview of the motivation behind and functions of private 5G networks~\cite{aijaz_private_2020}.
They present a large number of potential use cases and benefits for the IIoT and portray private 5G networking as the future of industrial networks.
Wen et al. come to a similar conclusion while pointing out the early stage at which private 5G networking currently is~\cite{wen_private_2022}.
Actual implementation and management are currently only feasible for large companies with a networking background.
In their paper covering critical mMTC, Pokhrel et al.~\cite{pokhrel_towards_2020}, argue that although URLLC is capable of providing excellent real-time behavior, the cost of doing so means relatively few devices can be connected in such a fashion.
Similarly, where mMTC is capable of connecting a plethora of devices, the individual QoS is not compatible with sensor networks.
To be able to meet the need for a large, wireless sensor network, they propose ``critical mMTC``.

\subsection{Device-Layer Technologies}
Only a few works have considered making improvements to the lowest layer in IIoT networks. The following research papers are concerned with securing embedded devices that are (somewhat newly) connected to large networks and responsible to execute real-time tasks.
Blumschein et al. present a real-time aware IP networking stack for IoT devices \cite{blumschein_differentiating_2022}. By classifying packets as early as possible and inheriting real-time task priorities to the networking task, they prevent priority inversions between a high-priority task and the processing of lower-priority packets. Additionally, incoming packet rates are limited to prevent a system overload from incoming packets.

In \cite{behnke_priority-aware_2022} a hardware-based approach with similar goals is presented. Using a multi-queue NIC design, packets are reordered and filtered based on the priority of the receiving real-time task. Furthermore, real-time aware interrupt moderation is proposed that reduces the number of interrupts generated by incoming traffic while guaranteeing low latencies for high-priority packets.
A SW/HW co-design for real-time aware packet reception is presented in \cite{behnke2023towards}.

Johansson et al. use MAC layer filtering to mitigate the effects of best-effort traffic reception on real-time tasks \cite{johansson2022priority}. They extend the network driver of VxWorks to support multiple receive queues and an interface that supports the configuration of the Ethernet Controller's MAC filter.
This way, packets are filtered and enqueued based on their priority in hardware, making it possible to treat incoming traffic based on real-time considerations.

In~\cite{austad_ftc}, TSN was used to extend Timed C, an extension to the C programming language, and include network channels as a primitive in the language.
When faced with large network loads, the critical traffic was reliably forwarded with minimal delay and jitter demonstrating the usefulness of TSN in distributed real-time systems.

\section{Discussion}
\label{sec:discussion}
Traditionally, industrial networks have been isolated, mostly homogeneous, and designed with dedicated protocols and hardware to meet stringent reliability and real-time requirements.
With the coming transformation of Industry 4.0 and IT/OT convergence, industrial networks need to become more configurable and flexible, enabling a wider range of applications than before.
This, of course, requires tighter connectivity between networked systems, as well as a higher degree of dynamism for nodes joining and leaving the network.

\subsection{Bridging the Gap between IIoT Scenarios and Network Requirements}
Looking at Table \ref{tbl:network_tech}, the wireless networking requirements of Scenario Classes I and VI appear to be the most difficult to meet.
While several wireless communication standards exist, they either cannot meet the strict latency predictability requirements or cannot meet the bandwidth requirements of multimedia data.
Once high reliability and latency are taken into account, current wireless networks do not scale sufficiently, i.e., a research gap exists where critical real-time requirements meet a need for wireless communication.
Current wireless real-time protocols are generally unable to scale to the required size and the aggregated bandwidth expected in the coming years.
Similarly, WLAN, and especially WLAN6E, can be seen as an attractive alternative with high bandwidth, low latency, improved timing accuracy, and readily available COTS hardware.

Comparing this to Table \ref{tab:comparison_recent}, it can be seen that this is where wireless TSN technologies will play a significant role in enablement.
The main gap-filler distilled from this work is wireless technologies that can provide a measure of predictability.
 However, the overall feasibility of this must be viewed with caution. While latency guarantees can be improved by real-time aware wireless packet switching, the inherent unpredictability of radio must not be overlooked.

While privately managed 5G networks seem to be a part of the solution, the high complexity and cost make it impractical for many companies and use cases.

The presented works regarding wired networks show that ``real-time networks`` are more static and carefully planned than traditional IT networks tend to be.
Even though technologies such as TSN help bring determinism to COTS Ethernet, achieving high flexibility and reconfigurability is a largely unsolved challenge.
Combining TSN with its centralized controllers (CNC, CUC) and SDN is one promising avenue that would extend to both WLAN and 5G solutions.
The downside of such a solution is the vast complexity of all systems; needless to say, the factory network of the future will be almost unrecognizable from today's, and be a major asset of its own.

Standardization and holistic large-scale implementations also are open challenges to the remaining subfields.
Even where research papers are feasible for specific scenario class requirements these only focus on individual aspects. 
The feasibility of the presented approaches still has to be validated in complex engineered systems.
Industry 4.0 is still missing a fixed set of standards and technologies which will be important for trust in safety and reliability. 
The approaches from the works presented have to be merged and implemented in holistic system architectures including all actors of the network, from embedded real-time devices, network devices, and compute nodes such as edge servers and cloud systems.

\section{Related Work}
\label{sec:related}
Several other works have looked at network requirements for industrial use cases. In the following, we present a selection of surveys from the past years that compare and analyze industrial networking solutions.

Xu et al.~\cite{xu2018survey} surveyed IIoT research from a systems perspective in 2018. In their work, they divide the wide research area into the three key aspects of control, networking, and computing before categorizing and investigating related works. While incorporating networking they do not cater to any real-time requirements. A more real-time focused survey was presented by Kim et al.~\cite{kim2017survey} in 2017.
In their work, the authors analyzed real-time communication in wireless sensor networks, which are in large part relevant to IIoT scenarios. Another survey on IIoT research was presented by Sisinni et al.~\cite{sisinni2018industrial} in 2018 with a focus on open challenges. Next to the need for energy efficiency, coexistence, interoperability, and security they also identify real-time performance as a future research challenge. 

Park et al. \cite{park_wireless_2018} presented requirements and challenges with using wireless networks in control applications. In particular, the focus lies on network design and relevant protocols.
Key benefits such as ease of installation and reduced operating costs are discussed. Several use cases, both relevant to current technologies (e.g. building automation, tire pressure sensors, wireless sensor coverage in industrial applications) and more futuristic examples where the avionics are brought out as an example where reducing cabling can provide 2-5\% weight reduction alone.
As wireless networks cannot provide the same isolation from interference as corresponding wired technologies, control loops must be adapted to tolerate packet loss and variable delays.
A joint design approach where control loops and the wireless network are jointly optimized to arrive at an acceptable performance is presented as one promising approach to improve the performance of wireless control systems.

In \cite{zunino_factory_2020}, Zunino et al. investigate the status of network technologies relevant to the Industrial Internet, or Industry 4.0.
They find that although most technologies needed to realize Industry 4.0 are present, they fall short on flexibility, scalability, reliability, real-time behavior, and security, to name a few.
A key point is to realize that "Industry 4.0" is a reaction to the change in market demands especially seen in developed countries.
A shorter time to market, a higher degree of customization, and small production batches are brought forward as key enablers to stay competitive.
To realize this, production lines must be highly configurable, have a high degree of situational awareness, and be extremely energy efficient.
Where current OT systems are tailored specifically for a particular task and hard real-time requirements are met by careful analysis and provisioning, future networks cannot be designed with such static scenarios in mind.
The communications networks must be able to provide real-time guarantees to a subset of nodes whilst also providing other QoS levels for various nodes, sensors, and systems.

Chen et al.~\cite{chen_wireless_2021} argue how smart manufacturing must use wireless technologies to orchestrate multi-robot systems to obtain the needed flexibility.
The third industrial revolution improved the flexibility of the \emph{quantity} produced, with the fourth, the flexibility of type, quantity, and quality being at the center of attention.
To change the production setup in such fundamental ways means that multi-robot systems must be highly configurable, the network must adapt to varying loads, and automated delivery systems like AGVs must be just as adaptable.

\section{Summary}
\label{sec:conclusion}
Real-time demands on networks in industrial environments have risen with the advent of the IoT paradigm and Industry 4.0. In this paper, we have listed common and anticipated use cases for industrial networks and identified the criteria needed to meet the requirements of each. 
We reviewed established industrial networking technologies and protocols and presented current academic works regarding the topic. 
By comparing different networking technologies to these criteria and use cases, we hope to give an overview of challenges and solutions concerning IIoT scenarios from a real-time networking perspective.

\addcontentsline{toc}{section}{References}
\printbibliography
\end{document}